\documentclass{epl}

\usepackage{amsmath}
\usepackage{amssymb}
\usepackage{calrsfs}

\newcommand{\ie}{\emph{i.}$\,$\emph{e.}}

\newcommand{\romd}{{\operatorname{d}}}

\newcommand{\VECe}{{\boldsymbol{e}}}
\newcommand{\VECf}{{\boldsymbol{f}}}
\newcommand{\VECl}{{\boldsymbol{l}}}
\newcommand{\VECn}{{\boldsymbol{n}}}
\newcommand{\VECt}{{\boldsymbol{t}}}
\newcommand{\VECx}{{\boldsymbol{x}}}

\newcommand{\VECF}{{\boldsymbol{F}}}
\newcommand{\VECX}{{\boldsymbol{X}}}

\newcommand{\VECnabla}{{\boldsymbol{\nabla}}}

\newcommand{\RR}{\mathbb{R}}

\title{Geometry of surface mediated interactions}
\shorttitle{Geometry of surface mediated interactions}

\author{Martin Michael M\"uller\inst{1} \and Markus Deserno\inst{1} \and Jemal Guven\inst{2}}
\shortauthor{M. M. M\"uller \etal} \institute{
\inst{1} Max-Planck-Institut f\"ur Polymerforschung, %
         Ackermannweg 10, %
         55128 Mainz, %
         Germany\\
\inst{2} Instituto de Ciencias Nucleares, %
         Universidad Nacional Aut\'onoma de M\'exico, %
         Apdo. Postal 70-543, %
         04510 M\'exico D.F., %
         Mexico}


\begin{document}

\maketitle


\begin{abstract}
  Soft interfaces can mediate interactions between particles bound to
  them.  The force transmitted through the surface geometry on a
  particle may be expressed as a closed line integral of the surface
  stress tensor around that particle. This contour may be deformed to
  exploit the symmetries present; for two identical particles, one
  obtains an exact expression for the force between them in terms of
  the local surface geometry of their mid-plane; in the case of a
  fluid membrane the sign of the interaction is often evident.  The
  approach, by construction, is adapted directly to the surface and is
  independent of its parameterization. Furthermore, it is applicable
  for arbitrarily large deformations; in particular, it remains valid
  beyond the linear small-gradient regime.
\end{abstract}


\section{Introduction}

Physical interactions between spatially separated particles are
mediated by fields: matter interacts by curving spacetime itself; if
it is charged it will interact through the electromagnetic field.
Equally well, interactions of an indirect nature may be mediated by
effective fields: for example, colloidal particles in suspension can
interact by disturbing local ion densities (if they are charged)
\cite{DLVO}, by distorting the order of an embedding liquid crystal
\cite{PoSt}, or by locally phase separating a binary mixture
\cite{binliq}.

An important class of such interactions, purely geometrical in nature,
are those between particles localized at an interface which are
induced through the local deformation in its shape. The simplest
example is provided by capillary interactions, originating in
\emph{surface tension} with an energy proportional to the excess area
of the interface. These interactions play a key role in many
technological processes, among them ore flotation and foam
stabilization \cite{colloid_and_flotation}.  To describe the
\emph{compression} and \emph{bending} of surfactant monolayers
and lipid membranes \cite{gbp,old_stress,inter_bend,weikl}, surface
energies involving higher order derivatives are required.  For
example, the interactions between membrane bound proteins, which play
an important role in cell biology \cite{Lodish}, are described by an
energy quadratic in the curvature of the membrane surface.

A major obstacle to providing a theoretical description of these
surface mediated interactions is that the corresponding field
equations are nonlinear.  It is easy to see why: while the underlying
surface free energy density may be simple --- some quadratic invariant
of generalized strains and thus ``harmonic'' --- the description of
the curved manifold it inhabits is intrinsically non-linear.  Even in
the case of an essentially flat surface, where its description in
terms of a height function (``Monge gauge'') keeping only the lowest
order in gradients leads to approximate linear differential equations
(and thus to Green functions), exact analytical solutions are
difficult to obtain. This is because a straightforward superposition
ansatz \cite{nicolson} will typically violate the correct boundary
conditions. It remains entirely unclear how any linear result
generalizes to the full nonlinear situation.

Despite this obstacle, substantial progress can still be made by
recognizing that the determination of the interaction between two
objects mediated by the surface involves solving two distinct
problems: first, the particular way in which objects bind to the
interface, together with its elastic specifications, determines its
equilibrium (\ie\ energy minimizing) shape---find that shape; second,
equilibrium of the resulting complex generally requires additional
external forces which constrain the bound objects at their
positions---find these forces.  Usually one thinks of the second
problem in a way which depends heavily on the successful
implementation of the first: since the energy of the complex depends
on the relative positions of the bound objects, appropriate
derivatives of the energy with respect to their coordinates yield the
forces one is looking for.  Rarely, however, is it possible to write
down this ``potential'' energy; as a result, this conceptually
straightforward route to the forces is infeasible in practice. In
other words, a solution to part 1 is generally impossible, leaving no
further handle on part 2.  This reasoning distracts from the fact that
forces transmitted by an interface must be encoded directly in its
shape; if the Hamiltonian describing the surface is geometrical, so
also are the stresses underpinning this shape.

In this letter we reformulate the full nonlinear interaction problem
in terms of surface geometry; the conserved covariant surface stress
tensor \cite{surfacestresstensor} associated with the geometry will
play a central role. We show how to express the surface mediated force
on a particle as a closed line integral of the stress tensor. By
suitably deforming the contour to exploit the symmetries of the
configuration, a remarkably simple and transparent expression for the
force is obtained. In particular, the \emph{sign} of the force may
turn out to be evident using only very qualitative features of the
geometry.  We should mention that a stress tensor approach to membrane
mediated interactions was suggested by Kralchevsky \etal\
\cite{old_stress}. However, the full potential of a completely
geometric description has not previously been exploited.


\section{Differential geometry, the energy, and the stress tensor}

In this section we sketch the essential geometric background.  While
this will define our notation, it is most likely too concise to serve
as a stand-alone introduction.  For the geometry we therefore refer
the reader to Refs.~\cite{DifferentialGeometry}; the stress tensor is
introduced at greater length in Ref.~\cite{surfacestresstensor,
Guven04}.

Consider a surface $\Sigma$ embedded in three-dimensional Euclidean
space $\RR^3$, which is described locally by its position
$\VECX(\xi^1,\xi^2)\in\RR^3$, where the $\xi^a$ are a suitable set of
(curvilinear) local coordinates on the surface.  Given the two tangent
vectors $\VECe_a=\partial \VECX/\partial\xi^a=\partial_a\VECX$ and the
(unit) normal vector $\VECn = \VECe_1\times\VECe_2 /
|\VECe_1\times\VECe_2|$, the surface geometry is described completely
in terms of the induced metric $g_{ab} = \VECe_a\cdot\VECe_b$ and the
extrinsic curvature $K_{ab}= -\VECn\cdot\partial_a\VECe_b$ \cite{Kab}.
We denote the metric-compatible covariant derivative by $\nabla_a$ and
the corresponding Laplacian as $\Delta=\nabla_a\nabla^a$.  As usual,
repeated indices -- one up one down -- imply a summation.  The total
curvature $K$ is the trace of the extrinsic curvature, $K =
g^{ab}K_{ab}$ \cite{GaussCodazziMainardi}.

We now associate with the surface an energy which can be written as a
surface integral over a scalar Hamiltonian density $\mathcal{H}$
constructed out of local geometric invariants,
\begin{equation}
  H_\Sigma[\VECX] = \int_\Sigma \romd A \;
  \mathcal{H}(g_{ab},K_{ab}, \nabla_aK_{bc}, \ldots) \ ,
  \label{eq:surfacefunctional}
\end{equation}
where the infinitesimal surface element is $\romd A = \romd^2 \xi
\sqrt{g}$ with $g=\det(g_{ab})$. Let us perform a variation $\VECX
\rightarrow \VECX + \delta\VECX$ of the embedding functions.  The
concomitant first variation of the functional
(\ref{eq:surfacefunctional}) can be cast as a bulk part plus a pure
divergence:
\begin{equation}
  \delta H_\Sigma =
  \int_\Sigma \romd A \; \mathcal{E}(\mathcal{H}) \, \VECn\cdot\delta\VECX
  \, + \,
  \int_\Sigma \romd A \; \nabla_aQ^a \ .
  \label{eq:deltaH_Sigma}
\end{equation}
Here, $\mathcal{E}\VECn$ is the \emph{bulk} Euler-Lagrange derivative
of $H_\Sigma$, which is evidently purely normal.  The second integral
(which is identical to the \emph{boundary integral} of $Q^a$ over
$\partial\Sigma$) originates in the tangential variations as well as
the derivatives of normal variations:
\begin{equation}
  Q^a = -\VECf^a\cdot\delta\VECX \; + \;
  \text{possible terms containing derivatives of } \delta\VECX \ .
  \label{eq:Qa}
\end{equation}
The object $\VECf^a$ is the \emph{surface stress tensor}.  Its
components in the local frame $\{\VECe_1,\VECe_2,\VECn\}$,
as well as the Euler Lagrange derivative $\mathcal{E}(\mathcal{H})$,
are listed for a few simple Hamiltonian densities in
Table~\ref{tab:surfacestresstensor}.

Suppose $\delta\VECX$ is simply a constant translation, which of
course leaves the Hamiltonian invariant.  We thus have $\delta
H_\Sigma=0$, and with the help of
Eqns.~(\ref{eq:deltaH_Sigma},\ref{eq:Qa}) $\nabla_a \VECf^a =
\mathcal{E}(\mathcal{H})\VECn$.  But a true equilibrium surface
is stationary with respect to \emph{arbitrary} variations. Thus
the Euler-Lagrange (``shape'') equation $\mathcal{E}=0$ also holds,
and we get the conservation law
\begin{equation}
  \nabla_a \VECf^a = 0 \ .
  \label{eq:divf=0}
\end{equation}
Its existence is simply a consequence of Noether's theorem: a
continuous symmetry implies a conservation law on shell.  Note that if
the surface encloses a fixed volume $V$, a term $-PV = -\frac{1}{3}P
\int_\Sigma \romd A \; \VECn\cdot\VECX$ involving the Lagrange
multiplier $P$ needs to be included in the functional, yielding the
shape equation $\mathcal{E}=P$, and Eqn.~(\ref{eq:divf=0}) is replaced
by $\nabla_a\VECf^a=P\VECn$.  The same holds if there exists a
pressure drop $P$ across the two sides of the interface.

\begin{table}
\begin{center}
\footnotesize
\begin{tabular}{cccc}
\hline\\[-0.6em]
  $\mathcal{H}$ & $\mathcal{E}$ & $f^a$ & $f^{ab}$ \\[0.3em]
\hline\\[-0.6em]
  $1$           & $K$ & $0$   & $-g^{ab}$ \\[0.8em]
  $K^n$         & $\big[R-\big(1-\frac{1}{n}\big)K^2-\Delta\big]nK^{n-1}$ & $-n\,\nabla^aK^{n-1}$ & $(nK^{ab}-Kg^{ab})K^{n-1}$ \\[1.2em]
  $K^{ab}K_{ab}$ & $\big[R-\frac{1}{2}K^2-\Delta\big]2K$ & $-2\,\nabla^a K$  & $(2K^{ab}-Kg^{ab})K$ \\[1.2em]
  $\begin{array}{c}\frac{1}{2}(\nabla K)^2\\ \equiv \frac{1}{2}(\nabla_cK)(\nabla^cK)\end{array}$ & \hspace*{-1em} $\begin{array}{c}(\Delta+K^2-R)\Delta K\\-K^{ab}\big[(\nabla_aK)(\nabla_bK)-\frac{1}{2}g_{ab}(\nabla K)^2\big]\end{array}$ \hspace*{-1em} & $\nabla^a\Delta K$ & $\begin{array}{c}(\nabla^aK)(\nabla^bK)\\-\frac{1}{2}g^{ab}(\nabla K)^2-K^{ab}\Delta K\end{array}$ \\[0.9em]
\hline
\end{tabular}
\end{center}
\caption{Euler-Lagrange derivative $\mathcal{E}(\mathcal{H})$
  and components of the stress tensor $\VECf^a = f^{ab}\VECe_b +
  f^a\VECn$ for several simple scalar surface Hamiltonian densities
  $\mathcal{H}$.  Projecting $\nabla_a \VECf^a = \mathcal{E}\VECn$
  implies the identities $\nabla_af^a-K_{ab}f^{ab}=\mathcal{E}$ and
  $\nabla_af^{ab}-K_a^bf^a=0$.  $R$ is the intrinsic scalar curvature
  \cite{GaussCodazziMainardi}.  Notice that $K^2$ and $K^{ab}K_{ab}$
  yield identical $\mathcal{E}$ and $\VECf^a$ (a consequence of the
  Gauss-Bonnet-Theorem \cite{DifferentialGeometry}).  These results 
  may be derived using techniques 
  developed in Refs.~\cite{surfacestresstensor,Guven04}.} \label{tab:surfacestresstensor}
\end{table}


\section{Forces via the stress tensor}

In elasticity theory the divergence of the stress tensor equals the
external force per unit volume of a strained material
\cite{LaLi_elast}.  Likewise, the divergence of the surface stress
tensor defined above equals the force per unit area of the strained
surface.  For instance, the equation $\nabla_a \VECf^a = P\VECn$ then
states that a pressure $P$ across the surface is a source of stress.
Using Stokes' theorem, the total force $\VECF_{\Sigma_0}$ acting
``within'' any patch $\Sigma_0$ is
\begin{equation}
  \VECF_{\Sigma_0} =
  \int_{\Sigma_0} \!\! \romd A \; \nabla_a\VECf^a =
  \oint_{\partial\Sigma_0} \!\!\! \romd s \; l_a\VECf^a \ ,
\end{equation}
where $\VECl=l^a\VECe_a$ is the \emph{outward} pointing unit normal to
the boundary curve $\partial\Sigma_0$ (which is by construction
tangential to the surface), and $s$ is the arc-length on
$\partial\Sigma_0$.  If $\Sigma_0$ is a \emph{free} equilibrium patch
(\ie, no external stresses), Eqn.~(\ref{eq:divf=0}) shows that
$\VECF_{\Sigma_0} = 0$.  Generally, however, the patch will contain
regions where external stresses act and $\VECF_{\Sigma_0}$ will be
nonzero.  Observe now that for $P=0$ the total force equals the line
integral of the stress tensor along \emph{any} curve enclosing these
sources.  This is because $\nabla_a\VECf^a=0$ permits us to deform the
contour of integration---provided we do not cross any of these sources
of stress in doing so. The case of $P\ne0$ will be treated elsewhere.

\begin{figure}
\onefigure[scale=1.00]{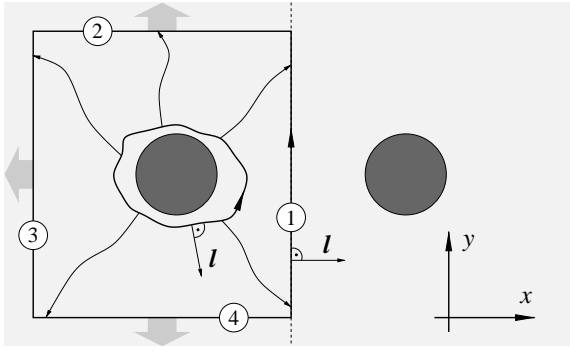}
  \caption{Illustration (view from the top) of how the force on one of
  a pair of objects bound to an interface can be calculated as a
  closed loop integral of the surface stress tensor. The contour 
  can subsequently be deformed to conform to the symmetry of the
  situation.}\label{fig:schematic}
\end{figure}

Consider now two identical particles bound to some asymptotically flat
surface, as is schematically sketched in Fig.~\ref{fig:schematic}. Due
to surface mediated interactions such a situation can only be
stationary if external constraining forces fix the particle
positions. These forces transmit stresses onto the surface, which are
thus picked up by a line integral over the surface stress tensor
around either particle.  In the absence of a pressure difference the
contour of the line integral can be deformed so as to take advantage
of the available symmetry, see again Fig.~\ref{fig:schematic}.  Once
the contour is pulled open wide enough, the surface will be flat at
branches 2, 3, and 4, and the stress tensor will be very simple.  In
fact, the contributions from branch 2 and 4 will then cancel each
other.  The only nontrivial contribution stems from branch 1, and its
evaluation is greatly simplified by the symmetry.

This outlines the basic strategy, which can evidently be tailored
towards many other situations.  We will now demonstrate its
application to a few important standard cases.


\section{Force in fluid membranes}

The example of a surface we would like to focus on is an elastic
symmetric fluid membrane, described by the surface Hamiltonian
\cite{Helfrich}
\begin{equation}
  \mathcal{H} = \frac{1}{2}\kappa \, K^2 + \sigma \ .
\end{equation}
Here, $\kappa$ is the bending stiffness and $\sigma$ the surface
tension.  For the special case $\kappa=0$ this reduces to the problem
of a surface with surface tension only, and describes a soap film on
large enough length scales or a water surface on length scales smaller
than the capillary length.  From Table~\ref{tab:surfacestresstensor}
we find that the associated stress tensor is given by
\begin{equation}
  \VECf^a
  \; = \;
  \Big[\kappa\big(K^{ab}-\frac{1}{2}Kg^{ab}\big)K - \sigma g^{ab}\Big]\,\VECe_b - \kappa(\nabla^aK)\,\VECn \ .
\end{equation}
We now introduce (orthonormal) tangent vectors $\{\VECt,\VECl\}$ along
branch 1: $\VECt$ points along the integration line ($\VECt = t^a
\VECe_a$), and (recall) $\VECl$ points normally outward.  A short
calculation then shows that the force stemming from branch 1 is given
by
\begin{equation}
  \VECF_1
  \; = \;
  -\int_1 \romd s \; \bigg\{
    \Big[\frac{1}{2}\kappa\big(K_\perp^2-K_{\|}^2\big)-\sigma\Big]\VECl
    -\kappa\big(\nabla_\perp K\big)\VECn\bigg\} \ ,
\end{equation}
where we have defined the principal curvatures $K_\perp = l^a l^b
K_{ab}$ and $K_{\|} = t^a t^b K_{ab}$, as well as the derivative along
$\VECl$, $\nabla_\perp= l^a \nabla_a = \partial/\partial l$.  By
symmetry there is no contribution to $\VECF_1$ along $\VECt$. The
minus sign out front stems from the fact that the membrane mediated
force on the particle is \emph{opposite} to the external force
necessary to counterbalance it.

\begin{figure}
\onefigure[scale=1.00]{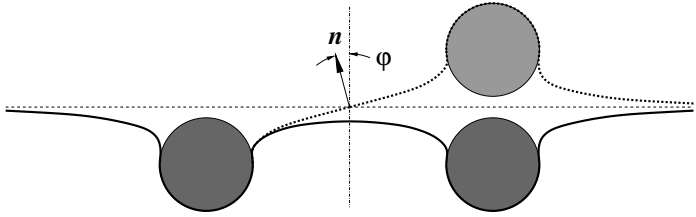}
  \caption{Illustration of the geometry of a symmetric (solid line)
  and an antisymmetric (dotted line) two-particle attachment.  The
  angle between surface normal at the symmetry point and vertical is
  $\varphi$.  }\label{fig:symmetry}
\end{figure}

To proceed further, we must look separately at the two different
possible symmetries: either two particles adhere at the same side of
the membrane (symmetric) or at different sides (antisymmetric), see
Fig.~(\ref{fig:symmetry}).  In the \emph{symmetric} case the
curvatures $K_\perp$ and $K_{\|}$ both have an extremum in the $\VECl$
direction along branch 1, so $\nabla_\perp K=0$ and the normal force
component vanishes there.  Furthermore, since the profile is
horizontal in the middle, $\VECl=\VECx$ there (where $\VECx$ is the
unit vector pointing in the horizontal $x$-direction). Finally, on
branch 3 the stress tensor simplifies to $\VECf_{a,3} =
-\sigma\VECe_a$, since far enough away the curvature becomes zero.
Therefore, the total force $\VECF_1 + \VECF_3 = F_{\text{sym}}\VECx$
on the left particle is given by
\begin{equation}
  F_{\text{sym}}
  \; = \;
  \sigma \Delta L -
    \frac{1}{2} \kappa \int_1 \romd s \; \big(K_\perp^2-K_{\|}^2\big) \ ,
\end{equation}
where $\Delta L>0$ is the excess length of branch 1 compared to branch
3.  We immediately see that the contribution due to tension is
attractive. The curvature contribution, on the other hand, is the
integral over the difference between the squared principal curvatures
along the mid-curve, and as such has no evident sign.  However, if we
had adsorbed two parallel \emph{cylinders}, which are sufficiently
long such as to neglect end effects, the contribution $K_{\|}^2$
vanishes. Furthermore, the mid-curve becomes a line, and thus $\Delta
L = 0$.  In this case we find for the force per unit length of the
cylinder
\begin{equation}
  F_{\text{sym,cyl}}/L
  \; = \;
  -\frac{1}{2} \kappa K_\perp^2
  \; \le \; 0 \ ,
  \label{eq:F.sym.cyl}
\end{equation}
which is evidently always repulsive.  Note that even though the
tension $\sigma$ does not occur explicitly, it will enter the force
indirectly through its influence on the value of $K_\perp$.  In Monge
gauge, with height function $h$, $K_\perp = [h'/(1+h'^2)]'|_{x=0} =
h''(0)$, since $h'(0)=0$.  Using this, Eqn.~(\ref{eq:F.sym.cyl}) is
then quantitatively corroborated (in the linear regime) by the
calculations in Ref.~\cite{weikl}.  Details of this will be presented
elsewhere.

In the \emph{antisymmetric} case branch 1 is a rotational symmetry
axis of degree 2 and thus a line; hence both $K_{\|}$ and $K_\perp$
vanish.  Since $K_\perp$ changes sign from positive to negative,
$\nabla_\perp K_\perp<0$.  Observe that the profile on the midline is
always tilted in the direction indicated in Fig.~\ref{fig:symmetry},
because otherwise it would have \emph{three} nodal points on the
asymptotic horizontal line and not just one, and the energy is
expected to be higher.  If we conceive of the constraining external
forces as fixing the \emph{horizontal separation} but not the vertical
position of the particles, the latter is equilibrated and the vertical
force component vanishes.  The force which remains on the left
particle is thus again horizontal, $\VECF_{\text{antisym}} =
F_{\text{antisym}}\VECx$, and given by
\begin{equation}
  F_{\text{antisym}}
  \; = \;
  \int_1\romd s \; \Big[\sigma\big(\cos\varphi(s)-1\big)
    - \kappa \sin\varphi(s)\,\nabla_\perp\big(K_\perp+K_{\|}\big)\Big] \ .
\end{equation}
This time the tension contribution is repulsive.  In the bending
contribution the $\nabla_\perp K_\perp$ is attractive, but
unfortunately not much can be said about the sign of $\nabla_\perp
K_{\|}$, hence the overall sign is not obvious.  However, in the case
of two cylinders bound on opposite sides, Eqn.~(\ref{eq:divf=0})
implies that $|\VECf^a|$ is constant on each of the three membrane
segments, and the force expression (per unit length) simplifies to
\cite{sign}
\begin{equation}
  F_{\text{antisym,cyl}}/L
  \; = \;
  |\VECf^\perp_{\text{midpoint}}| - \sigma
  \; = \;
  \sqrt{\sigma^2+(\kappa\nabla_\perp K_\perp)^2} - \sigma
  \; \ge \; 0 \ ,
  \label{eq:F.antisym.cyl}
\end{equation}
which is manifestly positive, implying particle attraction.  If we
expand the square root to first order in the bending part, we arrive
at the result for linear theory $F_{\text{antisym,cyl,$\varphi\ll
1$}}/L = \frac{1}{2}\kappa\,(\lambda\nabla_\perp K_\perp)^2$ (with
$\lambda=\sqrt{\kappa/\sigma}$).  This latter expression is again
quantitatively confirmed by using profile and force as they are
calculated in Ref.~\cite{weikl}.


\section{Discussion}

In the previous sections we have outlined a very general method to
obtain exact results for surface mediated interactions which sidesteps
the need to solve the field equations explicitly.  This approach is
fully covariant: one is free to choose the parameterization which is
most appropriate; it is also valid for large deformations and
\emph{not} limited to a linear approximation.  To our knowledge, there
are no analogues based on energy minimization of our results outside
this regime. It is true that our formulas contain unknown quantities
related to the shape of the surface.  It was clear from the beginning,
however, that we could not expect to solve the problem completely
without determining this shape; what has been established is the
connection between the \emph{geometry} of the surface and the
\emph{forces} transmitted by it, something which does not come across
when energy is differentiated -- even in the linear regime.  With
sufficient ingenuity, it may be possible to extract the necessary
information on the shape from the Euler-Lagrange equation (which so
far we have not used at all!) without needing to solve it
explicitly. Even without this input the functional form alone may
identify the sign of the interaction \cite{PoissonBoltzmann}.  Most
importantly, this framework provides a new window onto an old and
important problem. It can be combined with any approach, be it
analytical or numerical, which determines the surface shape. At the
very least, we have shown that it provides valuable non-trivial
consistency conditions for analytical calculations.  When such
calculations are ruled out, the determination of the force between
particles by numerically integrating the stress around one of them is
a process which is not only more straightforward but also considerably
more economical than calculating the energy as a function of distance
and numerically differentiating it.  A detailed description of the
interaction between two identical particles using this approach will
be presented elsewhere.


\acknowledgments

We have benefitted from discussions with Riccardo Capovilla.  We are
also grateful for the hospitality of the Dublin Institute for Advanced
Studies, where part of this work was completed.  MD acknowledges
financial support by the German Science Foundation through grant
De775/1-2.




\begin{thebibliography}{99}

\bibitem{DLVO}
\Name{Derjaguin B. V. \and Landau L. D.} \REVIEW{Acta
Physicochim. (USSR)}{14}{1941}{633}; \Name{Verwey E. J. \and
Overbeek J. T. G.} \Book{Theory of the stability of Lyophobic
  Colloids} (Elsevier, Amsterdam, 1948);
The analogy with capillarity is discussed in: \Name{Paunov V. N.}
\REVIEW{Langmuir}{14}{1998}{5088}.

\bibitem{PoSt}
\Name{Poulin P, Stark H, Lubensky T. C. \and Weitz D. A.}
\REVIEW{Science}{275}{1997}{1770}; \Name{Galatola P., Fournier
J.-B. \and Stark H.} \REVIEW{Phys. Rev. E}{67}{2003}{031404}.

\bibitem{binliq}
\Name{Fisher M. E. \and de Gennes P.-G.} \REVIEW{C. R. Acad. Sc.
Paris}{B287}{207}{1978}; \Name{Beysens D. \and Est{\`e}ve D.}
\REVIEW{Phys. Rev. Lett.}{54}{1985}{2123}.

\bibitem{colloid_and_flotation}
\Name{Russel W. N., Saville D. A. \and Schowalter W. R.}
\Book{Colloidal Dispersions} (Cambridge University Press,
Cambridge, U.K., 1989); \Name{Nguyen A. V. \and Schulz H. J.}
(eds.), \Book{Colloidal Science of Flotation} (Surfactant Science
Series \Vol{118}, Marcel Dekker, New York, 2003).

\bibitem{gbp}
\Name{Goulian M., Bruinsma R. \and Pincus P.} \REVIEW{Europhys.
Lett.}{22}{1993}{145}; Erratum: \REVIEW{Europhys.
Lett.}{23}{1993}{155}; note also the further correction in:
\Name{Fournier J. B. \and Dommersnes P. G.} \REVIEW{Europhys.
Lett.}{39}{1997}{681}.

\bibitem{old_stress}
\Name{Kralchevsky P. A., Paunov V. N., Denkov N. D. \and Nagayama
K.} \REVIEW{J. Chem. Soc. Farad. Trans.}{91}{1995}{3415};
\Name{Kralchevsky P. A. \and Nagayama K.} \REVIEW{Adv. Colloid
Interface Sci.}{85}{2000}{145};

\bibitem{inter_bend}
\Name{Kim K. S., Neu J. \and Oster G.} \REVIEW{Biophys.
J.}{75}{1998}{2274}; \Name{Rudnick J. \and Bruinsma R.}
\REVIEW{Biophys. J.}{76}{1999}{1725}; \Name{Marchenko V. I. \and
Misbah C.} \REVIEW{Eur. Phys. J. E}{8}{2002}{477}; \Name{Biscari
P., Bisi F. \and Rosso R.} \REVIEW{J. Math. Biol.}{45}{2002}{37};
\Name{Fournier J. B., Dommersnes P. G. \and Galatola P.}
\REVIEW{C. R. Biologies}{326}{2003}{467}.

\bibitem{weikl}
\Name{Weikl T. R.} \REVIEW{Eur. Phys. J. E}{12}{2003}{265}.

\bibitem{Lodish}
\Name{Lodish H., Zipursky S. L., Matsudaira P., Baltimore D. \and Darnell J.}
\Book{Molecular Cell Biology} (Freeman \& Company, New
  York, 2000).

\bibitem{nicolson}
\Name{Nicolson M. M.} \REVIEW{Proc. Cambridge Philos.
Soc.}{45}{1949}{288}.

\bibitem{surfacestresstensor}
\Name{Capovilla R. \and Guven J.} \REVIEW{J. Phys. A: Math.
Gen.}{35}{2002}{6233}.

\bibitem{DifferentialGeometry}
\Name{Kreyszig E.} \Book{Differential Geometry} (Dover, New York
1991); \Name{Do Carmo M.} \Book{Differential Geometry of Curves
and Surfaces} (Prentice Hall, 1976).

\bibitem{Guven04}
\Name{Guven J.} \REVIEW{J. Phys. A: Math. Gen.}{37}{2004}{L313}.

\bibitem{Kab}
Locally a smooth surface can be represented by a differentiable height
function $h(x,y)$ above the local tangent plane.  In these coordinates
$K_{ab}$ is simply the Hessian of $h$ divided by the square root of
the metric determinant: $K_{ab}=-\partial_a\partial_b h/\sqrt{g}$
($=-\partial_a\partial_b h$ to lowest order in $\VECnabla h$).

\bibitem{GaussCodazziMainardi}
We recall that the intrinsic and extrinsic geometries are related by
the Gauss-Codazzi-Mainardi equations $R_{abcd} = K_{ac}K_{bd} -
K_{ad}K_{bc}$ and $\nabla_aK_{bc} = \nabla_bK_{ac}$.  Here $R_{abcd}$
is the Riemann tensor, which quantifies the extent to which covariant
derivatives fail to commute: $[\nabla_a,\nabla_b]\VECe_c = R_{abcd}
\VECe^d$.  These equations occur as the
integrability conditions on the formulae of Gauss, $\nabla_a \VECe_b =
- K_{ab} \VECn$, and Weingarten, $\nabla_a \VECn = K_{ab}\VECe^b$,
obtained by taking a second derivative $\nabla_c$, switching the
indices $a$ and $c$ and subtracting \cite{DifferentialGeometry}. The
intrinsic scalar curvature $R = g^{ac}g^{bd}R_{abcd}$ thus satisfies
$R = K^2 - K^{ab}K_{ab} = 2\,K_{\text{G}}$, where
$K_{\text{G}}=\det(K_a^b)$ is the Gaussian curvature.

\bibitem{LaLi_elast}
\Name{Landau L. D. \and  Lifshitz E. M.} \Book{Theory of
Elasticity} (Butterworth-Heinemann, Oxford, 1999).

\bibitem{Helfrich}
\Name{Helfrich W.} \REVIEW{Z. Naturforsch.}{28c}{1973}{693};
\Name{Seifert U.} \REVIEW{Adv. Phys.}{46}{1997}{13}.

\bibitem{sign}
Expanding $\VECf^\perp$ in the local coordinate frame one finds
$\VECf^\perp\cdot\VECx = \text{sign}(\VECe_{\perp} \cdot
\VECx/f_{\perp})\,|\VECf^\perp|$.  Since $\VECe_{\perp} \cdot \VECx =
\cos\varphi>0$ and $f_{\perp,\text{midpoint}}=-\sigma<0$, we find
$\VECf^\perp_{\text{midpoint}}=-|\VECf^\perp|\VECx$, and thus
Eqn.~(\ref{eq:F.antisym.cyl}).

\bibitem{PoissonBoltzmann}
We mention as an aside that the same trick has recently rendered
possible a proof of the (nontrivial!) fact that within mean-field
electrostatics two identical like-charged colloids repel.  Here, the
Euler-Lagrange equation of the underlying field theory is the
nonlinear Poisson-Boltzmann equation, which cannot be solved
analytically in this geometry, but certain properties of the
corresponding (Maxwell) stress tensor nevertheless permit the
determination of the \emph{sign} of the interaction.  See:
\Name{Neu J.} \REVIEW{Phys. Rev. Lett.}{82}{1999}{1072};
\Name{Sader J. E. \and Chan D. Y. C.} \REVIEW{J. Colloid
Interface Sci.}{213}{1999}{268}; \Name{Trizac E.} \REVIEW{Phys.
Rev. E}{62}{2000}{R1465}.

\end{thebibliography}
\end{document}